\documentclass[aps,prl,twocolumn,showpacs]{revtex4}
\usepackage[dvips]{graphicx}
\usepackage{amsmath}
\usepackage{amsfonts}
\usepackage{amssymb}
\def\mywidth{0.8\columnwidth}
\begin{document}

\title{Distributed Delays Facilitate Amplitude Death of Coupled Oscillators}
\author{Fatihcan M. Atay}
\date{\emph{Physical Review Letters} {\bf 91}, 094101 (2003). DOI: 10.1103/PhysRevLett.91.094101} 

\affiliation
{Max Planck Institute for Mathematics in the Sciences,
Leipzig 04103, Germany}
\email{atay@member.ams.org}
\homepage{http://private-pages.mis.mpg.de/fatay}
\pacs{02.30.Ks, 05.45.Xt, 87.10.+e \hfill \textbf{Copyright:} \emph{American Physical Society} }%

\begin{abstract}%
Coupled oscillators are shown to experience amplitude death for a much larger
set of parameter values when they are connected with time delays distributed
over an interval rather than concentrated at a point. Distributed delays
enlarge and merge death islands in the parameter space. Furthermore, when the
variance of the distribution is larger than a threshold the death region
becomes unbounded and amplitude death can occur for any average value of
delay. These phenomena are observed even with a small spread of delays, for
different distribution functions, and an arbitrary number of oscillators.
(\href{http://link.aps.org/abstract/PRL/v91/e094101}{http://link.aps.org/abstract/PRL/v91/e094101})

\vspace{0.2cm}

\end{abstract}%
\maketitle

Coupled oscillators constitute an effective and popular paradigm for the study
of interacting oscillatory processes in the physical and biological sciences
\cite{Linkens77}.
The rich dynamics arising from the interaction
of simple units have been a source of interest for scientists modeling the
collective behavior of real-life systems. Among the most widely studied
phenomena is synchronization, where individual units oscillate at a common
frequency and phase when coupled \cite{Pikovsky-book01}.  
Synchronization may be observed even under weak coupling;
so it has usually been studied through reduced models that retain only the
phase information along the limit cycles. With stronger coupling further
interesting behavior is possible, whose investigation requires the use of full
models that include the amplitudes of the oscillators. An example is
amplitude death, which refers to the quenching of oscillations under
coupling, as the system evolves to a stable equilibrium
\cite{Yamaguchi84,Bar-Eli85}. If the information flow through the coupled
system is instantaneous, amplitude death occurs when 
the individual oscillators have sufficiently different frequencies.
\cite{Aronson90,Ermentrout90,Mirollo90}.
On the other hand, if the information
from one oscillator reaches the others after a certain time delay, which may
be due to finite propagation or information processing speeds, then even
identical oscillators can experience amplitude death when coupled
\cite{Reddy98}. Recent experimental and theoretical studies have
confirmed the role of delays in inducing amplitude death
\cite{Herrero00,Reddy00b,Atay-PHYSD03}.

While the importance of time delays in amplitude death is now clear, 
studies in this area have so far been confined
only to discrete, or constant, delays. In other words, it has been assumed
that information reaches from one unit to another after a fixed time $\tau$
which is unchanging as the system evolves, and moreover, the
units act only on the instantaneous value of the received information
and forget any previous values. Such discrete-delay models often fail to
adequately describe physical systems by neglecting the possibilities that (a)
the quantity $\tau$ may only be approximately known, (b) it may only represent
an average value of a quantity that varies between pairs of oscillators in a
network or (c) varies in time through a process involving unmodelled factors,
and (d) the oscillators may incorporate ``memory" effects by using the past
history of the received information. The first possibility is certainly an
issue in any experimental situation, (b) is typical when considering large
arrays, and (c) and (d) are particularly significant in biology and neurology.
Because of these shortcomings, models based on \emph{distributed} delays have
been proposed as early as the time of Volterra \cite{Volterra31}, and used in
such areas as biology \cite{MacDonald78}, ecology
\cite{May74,Gopalsamy92}, neurology \cite{Baylor74}, viscoelasticity
\cite{Drozdov-Kolmanovskii}, and economics \cite{Haldane33}. It has especially
been pointed out in the biological sciences that distributed delays lead to
more realistic models \cite{Cushing77}. In this Letter we consider the effects
of distributed delays on amplitude death, and show
that even a small spread in the delay distribution can greatly enlarge the set
of parameters for which amplitude death occurs.

The oscillators studied here are described by
\begin{equation}
\dot{Z}(t)=(1+\mathrm{i}\omega_{0}-|Z(t)|^{2})Z(t), \label{single}%
\end{equation}
where $Z(t)$ is a complex number and $\omega_{0}>0$. Equation (\ref{single})
represents the normal form for a supercritical Hopf bifurcation, and has been
used to describe limit cycle oscillators where oscillations arise through such
a bifurcation. It has an unstable equilibrium at zero, and an attracting limit
cycle $Z(t)=\exp(\mathrm{i}\omega_{0} t)$ with frequency $\omega_{0}%
$. Suppose a pair of such oscillators are coupled with general time delays:
\begin{align}
\dot{Z}_{1}(t)  &  =(1+\mathrm{i}\omega_{1}-|Z_{1}(t)|^{2})Z_{1}(t)\nonumber\\
&  +K\left[  \int_{0}^{\infty}f(\tau^{\prime})Z_{2}(t-\tau^{\prime})
d\tau^{\prime}-Z_{1}(t)\right]  \label{coup1}%
\end{align}%
\begin{align}
\dot{Z}_{2}(t)  &  =(1+\mathrm{i}\omega_{2}-|Z_{2}(t)|^{2})Z_{2}(t)\nonumber\\
&  +K\left[  \int_{0}^{\infty}f(\tau^{\prime})Z_{1}(t-\tau^{\prime})
d\tau^{\prime}-Z_{2}(t)\right]  . \label{coup2}%
\end{align}
Here, $K$ is a number quantifying the strength of coupling, and $f$ represents
a distribution of delay values. When $f$ is the delta function
$\delta(0)$ one obtains the system considered in \cite{Aronson90}, where the
oscillators interact without delay. Similarly, the choice $\delta(\tau)$ with
$\tau>0$ gives the system with a discrete delay which was studied in
\cite{Reddy98}. In the general case $f$ is a probability density over an
appropriate interval, which addresses the shortcomings mentioned in the above
paragraph.
We shall show that the variance of $f$ has a significant effect 
on the dynamics of the system.

When the system (\ref{coup1})--(\ref{coup2}) experiences amplitude
death its zero solution becomes stable. To investigate the stability, the
system is linearized about zero. The characteristic equation
is found by making the ansatz $(Z_{1}(t),Z_{2}(t))=\xi\exp(\lambda
t)$, $\xi\in\mathbf{R}^{2}$, and is given by
\begin{equation}
(1+\mathrm{i}\omega_{1}-K-\lambda)(1+\mathrm{i}\omega_{2}-K-\lambda
)-K^{2}\left[  F(\lambda)\right]  ^{2}=0 \label{char}%
\end{equation}
where $F$ is the Laplace transform of $f$. For definiteness, the analysis of
(\ref{char}) will be illustrated for uniformly distributed delays
over the interval $\tau\pm\alpha$, i.e.
$f(\tau^{\prime})=1/(2\alpha)$ if $|\tau-\tau^{\prime}|\leq\alpha$ and zero
 otherwise.
However, the idea is the same for other types of distributions.
We first consider oscillators with identical
frequencies, $\omega_{1}=\omega_{2}=\omega_{0}$. This is the more stringent
case for stability, because different frequencies can
stabilize coupled oscillators even in the absence of delays
whereas identical frequencies cannot
\cite{Aronson90,Ermentrout90}. The analysis is based on the
observation that as parameters are varied the stability of the origin may
change only if an eigenvalue $\lambda$ crosses the imaginary axis. In this
critical situation $\lambda=\mathrm{i}\omega$ for some real $\omega$, and
by (\ref{char})
\begin{equation}
(1+\mathrm{i}(\omega_{0}-\omega)-K)^{2}-K^{2}\gamma^{2}e^{-2\mathrm{i}%
\omega\tau}=0 \label{char-i}%
\end{equation}
where
\begin{equation}
\gamma=\gamma(\omega,\alpha)=\left\{
\begin{array}
[c]{ccc}%
\sin(\omega\alpha)/(\omega\alpha) & \text{if} & \omega\alpha\neq0,\\
1 & \text{if} & \omega\alpha=0.
\end{array}
\right.  \label{gamma}%
\end{equation}
Separating (\ref{char-i}) into real and imaginary parts and rearranging
yields
\begin{align}
(1-\gamma^{2})K^{2}-2K+1  &  =-(\omega-\omega_{0})^{2}\label{curve1}\\
\tan(\omega\tau)  &  =\frac{\omega-\omega_{0}}{1-K}. \label{curve2}%
\end{align}
This pair of equations describe a set of parametric curves on the $\tau$-$K$
plane in the parameter $\omega$. For each value of $\omega$, $K$
is found from (\ref{curve1}), and substitution into
(\ref{curve2}) gives the corresponding values for $\tau$.
The stability region is determined by computing the
critical curves (\ref{curve1})--(\ref{curve2}), and following the direction of
movement of the purely imaginary eigenvalues as parameters are varied. The
latter information is obtained from the quantities $\operatorname{Re}%
(\partial\lambda/\partial K)$ and $\operatorname{Re}(\partial\lambda
/\partial\tau)$ calculated from (\ref{char}) by implicit differentiation on
the critical curves. For the parameter values obtained by this procedure,
amplitude death is independently confirmed by numerical simulation of the
coupled system (\ref{coup1})--(\ref{coup2}).

\begin{figure}[tb]
\includegraphics[width=\columnwidth]{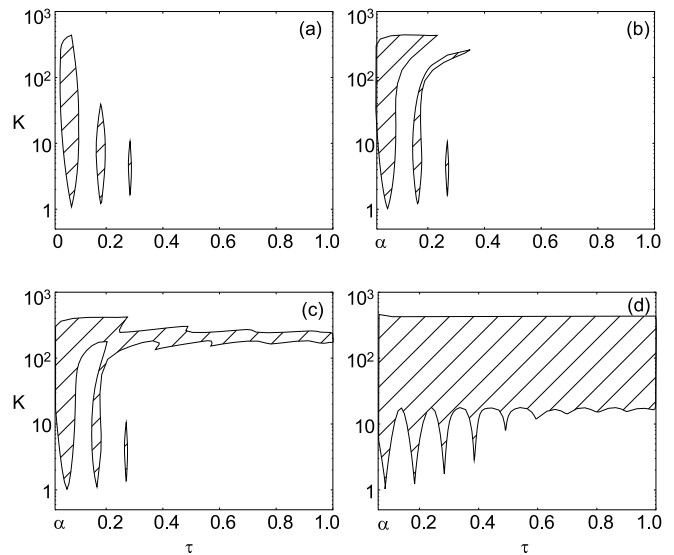}
\caption{Enlargement of the stability
region as the spread of the delay distribution increases: (a) $\alpha=0$, (b)
$\alpha=0.007$, (c) $\alpha=0.008$, and (d) $\alpha=0.02$. The value of
$\omega_{0}$ is 30. }%
\label{fig:K-tau}%
\end{figure}

The stability region in the $K$-$\tau$ parameter plane is shown in
Fig.~\ref{fig:K-tau} for uniformly distributed delays over $\tau\pm\alpha$,
with $\tau\geq\alpha$ so that the delays involved are not negative.
As $\alpha\rightarrow0$, the uniform distribution approaches
$\delta(\tau)$; hence $\alpha$ serves as a parameter to compare the
discrete and the uniformly distributed delays having the same mean value
$\tau$. This is equivalent to quantifying the effects of the
distribution by its standard deviation $\sigma=\alpha/\sqrt{3}$.
When the delay is discrete the
stability region consists of three disjoint and bounded sets
(Fig.~\ref{fig:K-tau}a) which deform continuously as $\alpha$ is increased
from zero (Fig.~\ref{fig:K-tau}b). The enclosed area increases
with the variance of the distribution, and at some critical value the
stability region becomes unbounded in the $\tau$-direction
(Fig.~\ref{fig:K-tau}c). The critical value of $\alpha$ for this qualitative
transition depends on $\omega_{0}$ and it can be very small; it is slightly
below 0.008 when $\omega_{0}=30$, corresponding to
$\sigma=4.62\times10^{-3}$. Increasing $\alpha$ further results in a
connected and unbounded set of parameter values for amplitude death
(Fig.~\ref{fig:K-tau}d), which persists for all larger values of $\alpha$. At
this stage, there is a large interval of values for the coupling strength $K$
which causes amplitude death regardless of the mean value $\tau$. By contrast,
the discrete delay can cause amplitude death only for
a very limited range of delay values (Fig.~\ref{fig:K-tau}a).
Note that the ratio $\sigma/\tau$ of the
standard deviation to the mean of the distribution can be quite small, showing
that even a relatively small spread in delays may induce amplitude death.

\begin{figure}[tb]
\includegraphics[width=\mywidth]{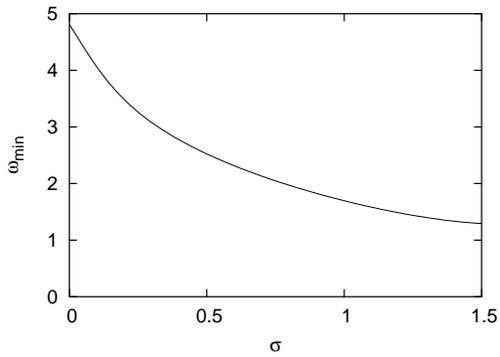}
\caption{Minimum value of $\omega_{0}$
for which amplitude death is possible decreases as the spread of 
the delay distribution increases.}%
\label{fig:minw0}%
\end{figure}

Distributed delays further facilitate amplitude death through the parameter
$\omega_{0}$. The stability region in the parameter plane depends on
the value of $\omega_{0}$, and there exists a minimum value $\omega_{min}$
such that if $\omega_{0}<\omega_{min}$ then amplitude death does not occur for
any $K$ or $\tau$. %A numerically calculated
A value of
$\omega_{min}=4.812$ has been reported for discrete delays \cite{Reddy98}.
Distributed delays can induce amplitude death even when $\omega_{0}$
is smaller than this value, as Fig.~\ref{fig:minw0} shows. 
The curve has been numerically calculated by decreasing the value of 
$\omega_0$ at a fixed $\sigma$ for the uniform distribution until 
the stability region disappears.

\begin{figure}[tb]
\includegraphics[width=\mywidth]{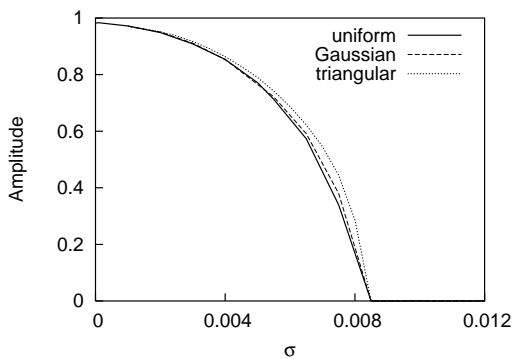}
\caption{Amplitudes of the 
limit cycles of individual oscillators versus the standard deviation of 
the delay distribution, for common distribution functions with the 
same mean value $\tau=0.5$. Other parameters are $K=30$ and 
$\omega_{0}=30$.}%
\label{fig:amplitude}%
\end{figure}

The local stability change caused by distributed delays is reflected in
the global dynamics by the annihilation of the limit cycles of the
oscillators. Fig.~\ref{fig:amplitude} shows the %numerically calculated
amplitudes of the oscillators, which gradually decrease with increasing
standard deviation of the delay distribution. Near $\sigma=0.0085$
the amplitude becomes zero as the limit cycle collapses to the
origin and the amplitude death sets in.
Numerical simulations with random initial conditions indicate that for
$\sigma$ beyond this value the origin is the only attractor for the coupled
system. Notably, this behavior is largely independent
of the particular distribution chosen for the delays,
as seen in Fig.~\ref{fig:amplitude}.

\begin{figure}[tb]
\includegraphics[width=\columnwidth]{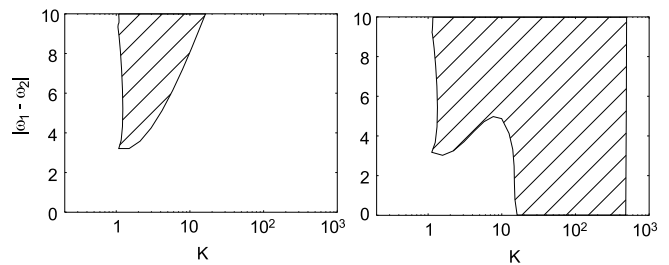} 
\caption{Stability
regions for coupled oscillators with different frequencies when the delay is
(a) discrete at $\tau=0.5$, and (b) uniformly distributed over $0.5\pm 0.02$. 
The mean frequency of the oscillators is fixed at 30.}%
\label{fig:freq}%
\end{figure}

When the oscillators have different intrinsic frequencies the situation is
similar, except that amplitude death occurs for even a larger parameter set,
since a large frequency difference by itself is known to cause death.
Fig.~\ref{fig:freq} compares the stability regions for discrete and distributed
delays. As the spread of the delays increase, the stability region is enlarged
and extended towards the horizontal axis, so amplitude death becomes possible
also for a small (or zero) frequency difference.

\begin{figure}[tb]
\includegraphics[width=0.68\columnwidth]{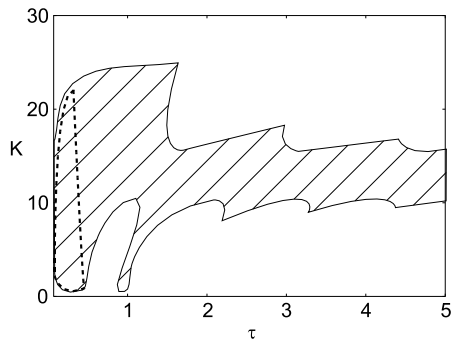}
\caption{The limiting shape of the stability
region for an infinite number of globally coupled oscillators, calculated for
delays uniformly distributed over $\tau\pm0.07$. The area enclosed by the
dashed line is the stability region for the discrete delay.}%
\label{fig:coupN}%
\end{figure}

The stabilizing effects of distributed delays carry over to any number of
oscillators. This is illustrated in the following system of $N$ globally
coupled oscillators:%
\begin{align}
\dot{Z}_{j}(t)  &  =(1+\mathrm{i}\omega_{j}-|Z_{j}(t)|^{2})Z_{j}(t)\nonumber\\
&  +\frac{2K}{N}\sum_{\substack{k=1\\k\neq j}}^{N}\left[  \int_{0}^{\infty
}f(\tau^{\prime})Z_{k}(t-\tau^{\prime})~d\tau^{\prime}-Z_{j}(t)\right]
\label{coupN}%
\end{align}
where $j=1,\dots,N$. In this model the use of distributed delays is
further motivated by the fact that in a physical network the delays between
units cannot be expected to be identical but can be more naturally
described by a probability distribution for large $N$. For
$N=2$, (\ref{coupN}) reduces to (\ref{coup1})--(\ref{coup2}). The system
(\ref{coupN}) was studied in \cite{Ermentrout90,Mirollo90} with no delays, and
in \cite{Reddy98} with discrete delays. For identical oscillators an analysis
similar to above yields the region of amplitude death for (\ref{coupN}). The
limiting shape of this region as $N\rightarrow\infty$ is depicted in
Fig.~\ref{fig:coupN} for uniformly distributed delays
and $\omega_{j}=10$ for all $j$. As before, distributed delays 
enlarge the stability region, and there exist values of $K$ for
which amplitude death occurs regardless of the mean value $\tau$ of the
delays. (The stability region is enlarged further if the frequencies $\omega_j$ 
are not identical, similar to the case shown in Fig.~\ref{fig:freq}.)
It is interesting to compare the results to those in
\cite{Ermentrout90,Mirollo90}, where it was shown that a sufficiently large
spread in the frequencies $\omega_{j}$ can cause amplitude death.
Here a similar conclusion holds for a spread in the delays.

A detailed mathematical analysis of stability under distributed delays
is too lengthy to include here; 
however, a brief description of the basic ideas will help clarify the role of
delay distributions. Thus consider the characteristic equation
(\ref{char}). If $F\equiv0$ the eigenvalues are $\lambda=1-K+\mathrm{i}%
\omega_{j},$ $j=1,2$, so the system is asymptotically stable for $K>1$. It
follows that stability can also be achieved for $|F|$ sufficiently small.
Letting $\lambda=\beta+\mathrm{i}\omega$,
\[
F(\lambda)=\int_{0}^{\infty}f(\tau^{\prime})e^{-\beta\tau^{\prime}}[\cos
\omega\tau^{\prime}-\mathrm{i}\sin\omega\tau^{\prime}]\,d\tau^{\prime}
\]
which can be interpreted as a (weighted) average of the quantity in brackets
over an interval determined by $f$. If $\omega\neq0$ the integrand
is oscillatory about zero, and smaller values of $|F|$ may be obtained if the
average is taken over a larger interval, i.e.~if $f$ has a large
variance. It can be assured that $\omega$ is bounded away from zero provided
$K$ is not too large. In this way, an interval of values of $K$ is obtained
for which the origin is stable. A rigorous argument involves assuming
$\beta\geq0$ and obtaining a contradiction resulting from $|F|$ being small.
It is also intuitively plausible that if $f$ has a
sufficiently large variance, then $|F|$ is small regardless of the mean value
or the precise shape of $f$. This explains Figures~\ref{fig:K-tau}d and
\ref{fig:coupN}, which show stability for a range of coupling strengths and
arbitrary mean value of delay, as well as Fig.~\ref{fig:amplitude}, 
which shows similar behavior for different distribution functions.

The importance of amplitude death has been noted by many authors in relation
to various physical and biological phenomena, ranging from
Belousov-Zhabotinskii reactions to cardiac arrhythmias
\cite{Bar-Eli85,Mirollo90,Reddy98}. For instance, the cessation of
rhythmic activity in biological systems may be related to certain pathologies.
Our findings suggest that in certain cases the variance of the delays, rather
than their average value, could be the relevant quantity responsible for the
quenching of oscillations. Delay distributions has a stabilizing effect on the
interconnected system, similar to that of frequency distributions treated in
previous works \cite{Ermentrout90,Mirollo90}. This implies that amplitude
death is a rather common and robust dynamical behavior for interacting
oscillatory processes, since real-life networks inevitably involve variances
in both the frequencies and the connection delays. From a practical point, the
properties of distributed delays are expected to be helpful in modelling
observed phenomena. For instance, a situation where amplitude death is
experimentally observed but not predicted by an existing model might imply the
presence of distributed delays in the physical system. 
Introducing a small variance in delays may in many
cases lead to a better reconciliation of theory with experiments.
Finally, distributed delays give a
natural way to model memory effects in interacting systems. They are thus
particularly suitable in fields of neuroscience, cognition, and
the general analysis of complex systems. Delays are expected to
be a source of further interesting results in these 
highly active areas of investigation.

\end{document}